\documentclass[10pt,journal,twocolumn]{IEEEtran}
\usepackage{ifpdf}
 \ifpdf
 \else
 \fi

\ifCLASSINFOpdf
   \usepackage[pdftex]{graphicx}
\else
   \usepackage[dvips]{graphicx}
\fi
\usepackage[cmex10]{amsmath}

\usepackage[tight,footnotesize]{subfigure}
\usepackage{define-notations,amssymb}
\usepackage{algorithmic,algorithm,cite,color}

\begin{document}
\title{Power minimization for OFDM Transmission with
Subcarrier-pair based Opportunistic DF Relaying}

\author{Tao Wang,~\IEEEmembership{Senior Member, IEEE}, Yong Fang,~\IEEEmembership{Senior Member, IEEE}
and Luc Vandendorpe,~\IEEEmembership{Fellow, IEEE}
\thanks{
T. Wang is with School of Communication \& Information Engineering,
Shanghai University, 200072 Shanghai, P. R. China.
He was with ICTEAM Institute, Universit\'e Catholique de Louvain (UCL),
1348 Louvain-la-Neuve, Belgium.
(email: t.wang@ieee.org).

Y. Fang is with School of Communication \& Information Engineering, Shanghai University, 200072 Shanghai, P. R. China
(email: yfang@staff.shu.edu.cn).

L. Vandendorpe is with ICTEAM Institute, UCL, 1348 Louvain-la-Neuve, Belgium
(email: luc.vandendorpe@uclouvain.be).

Research supported by The Program for Professor of
Special Appointment (Eastern Scholar) at Shanghai Institutions of Higher Learning.
It is also supported by the IAP project BESTCOM,
the ARC SCOOP, the NEWCOM\# and NSF China \#61271213.
}}


\maketitle

\begin{abstract}
This paper develops a sum-power minimized resource allocation (RA) algorithm
subject to a sum-rate constraint for cooperative
orthogonal frequency division modulation (OFDM) transmission
with subcarrier-pair based opportunistic decode-and-forward (DF) relaying.
The improved DF protocol first proposed in \cite{Vandendorpe08-2} is used
with optimized subcarrier pairing.
Instrumental to the RA algorithm design is appropriate definition
of variables to represent source/relay power allocation,
subcarrier pairing and transmission-mode selection elegantly,
so that after continuous relaxation, the dual method
and the Hungarian algorithm can be used to find
an (at least approximately) optimum RA with polynomial complexity.
Moreover, the bisection method is used to speed up the search
of the optimum Lagrange multiplier for the dual method.
Numerical results are shown to illustrate
the power-reduction benefit of the improved DF protocol
with optimized subcarrier pairing.
\end{abstract}

\begin{IEEEkeywords}
Cooperative communication, resource allocation, decode and forward, OFDM.

\end{IEEEkeywords}

\section{Introduction}

Cooperative orthogonal frequency division modulation (OFDM)
transmission with subcarrier-pair based decode-and-forward (DF)
relaying and associated resource allocation (RA) were studied in
\cite{WangYing07,Li08,Haj11,Vandendorpe08-1}
when the source-to-destination link exists.
In \cite{WangYing07,Li08,Haj11}, an ``always-relaying" DF protocol was considered.
To better exploit the frequency-selective channels,
we have proposed an opportunistic DF relaying protocol in \cite{Vandendorpe08-1},
i.e, a subcarrier in the first time slot can either be
paired with a subcarrier in the second slot for the relay-aided transmission,
or used for the direct source-to-destination transmission without the relay's assistance.
A major drawback for that DF protocol is that,
a subcarrier unused for relaying in the second slot becomes idle,
which wastes spectrum resource.
To address this issue, we first proposed in \cite{Vandendorpe08-2}
an improved DF protocol, which allows the source to make
the direct transmission over the subcarriers unused for relaying in the second slot.
This protocol and its RA were later intensively investigated,
e.g., in \cite{Vandendorpe09-1,Vandendorpe09-2,WangTSP11,WangJSAC11,Hsu10,Boost10}.
Note that in \cite{Vandendorpe08-1,Vandendorpe08-2,Vandendorpe09-1,Vandendorpe09-2,
WangTSP11,WangJSAC11}, a priori subcarrier pairing was used.
i.e., when the relay-aided transmission is used for a subcarrier in the first slot,
the same subcarrier in the second slot is always paired with
this subcarrier for DF relaying.
In \cite{WangYing07,Li08,Hsu10,Boost10,Haj11},
the optimization of subcarrier pairing was considered.

Recently, energy-efficient communication is becoming increasingly important \cite{Li11}.
In view of the fact that many existing works focus on spectral-efficiency maximized RA,
we develop an RA algorithm for minimizing the sum power
subject to a sum-rate constraint for the improved DF protocol
with optimized subcarrier pairing (OSP).
Compared with the algorithms designed in \cite{Haj11,Hsu10},
our algorithm uses a new method to define indicator variables
for representing subcarrier pairing and transmission-mode selection,
by regarding the subcarriers for the direct transmission
in the two time slots as virtual subcarrier pairs.
Moreover, the bisection method is used to find the optimum
Lagrange multiplier, which is faster than the incremental-update
based subgradient method used in \cite{Haj11,Hsu10}.

Notations: $\Rate{x} = \frac{1}{2}\log_2(1 + x)$ and $[x]^+ = \max\{x,0\}$.

\section{System and protocol description}

Consider the scenario where a relay assists a source's transmission 
to a destination. 
The improved DF protocol in \cite{Vandendorpe08-2} is used.
Specifically, every data-transmission session takes two consecutive
equal-duration time slots 
and OFDM with $K$ subcarriers is used.
To facilitate description, a subcarrier used in the first slot
is denoted by subcarrier $k$ and one in the second slot by subcarrier $l$ hereafter.
In the first time slot, the source radiates OFDM symbols,
using $\Pskone$ as the transmit power for subcarrier $k$.
The source-to-relay and source-to-destination baseband-channel
coefficients for subcarrier $k$ are $\hsrk$ and $\hsdk$, respectively.
In the second slot, both the source and the relay synchronously radiate OFDM symbols,
using $\Psltwo$ and $\Prl$ as the transmit powers for subcarrier $l$, respectively.
The relay-to-destination baseband-channel coefficient is $\hrdl$ for subcarrier $l$.

A subcarrier in the first slot can either be paired with one in the second slot
for the relay-aided transmission, or be used for the direct transmission without relaying.
Every unpaired subcarrier in the second slot is used for the direct transmission.
In particular, if subcarrier $l$ is used for the direct transmission,
$\Psltwo\geq 0$ is used while $\Prl=0$ is imposed.
The maximum average data rates over subcarriers $k$ and $l$
used for the direct transmission are $\Rate{\Pskone\Gsdk}$
and $\Rate{\Psltwo\Gsdl}$ bits/OFDM-symbol (bpos), respectively,
where $\Gsdk = \frac{|\hsdk|^2}{\sigma^2}$
and $\sigma^2$ is the noise variance for each subcarrier at every node's receiver.
When subcarrier $k$ is paired with subcarrier $l$
for the relay-aided transmission, the DF relaying is used
in which case $\Psltwo=0$ is imposed while $\Pskone\geq 0$
and $\Prl\geq 0$ are used (more details are available in \cite{Vandendorpe08-2}).
Suppose $\Pskone + \Prl = P$, it can readily be shown that
the maximum data rate is equal to $R_{k,l} = \Rate{\Gkl P}$ bpos, where
\begin{align}
\Gkl = \left\{\begin{array}{ll}
               \frac{\Gsrk\Grdl}{\Gsrk - \Gsdk + \Grdl}  & {\rm if\;}\min\{\Gsrk,\Grdl\}> \Gsdk,  \\
               \min\{\Gsrk,\Gsdk\}                  & {\rm if\;}\min\{\Gsrk,\Grdl\}\leq \Gsdk,
           \end{array}\right.  \nonumber
\end{align}
$\Gsrk = \frac{|\hsrk|^2}{\sigma^2}$ and $\Grdl = \frac{|\hrdl|^2}{\sigma^2}$
\cite{Vandendorpe08-2}.
This maximum rate is achieved when
\begin{align}
\Pskone = \left\{\begin{array}{ll}
               \frac{\Grdl}{\Gsrk - \Gsdk + \Grdl}P   & {\rm if\;}\min\{\Gsrk,\Grdl\}> \Gsdk,  \\
               P                                 & {\rm if\;}\min\{\Gsrk,\Grdl\}\leq \Gsdk,
           \end{array}\right. \nonumber
\end{align}

Assume there exists a central control unit which knows precisely
$\{\Gsrk,\Gsdk|\forall\;k\}$ and $\{\Grdl|\forall\;l\}$,
and determines the optimum RA (i.e., the source/relay power allocation, 
subcarrier pairing and transmission mode selection ) 
to minimize the sum power subject to the constraint that 
the sum data rate is not smaller than prescribed $\Rreq$ bpos.



\section{RA algorithm design}

For any subcarrier assignment used by the improved DF protocol,
suppose $m$ subcarrier pairs are assigned to the relay-aided transmission,
then it is always possible to one-to-one associate the unpaired subcarriers
in the two slots to form $K-m$ virtual subcarrier pairs for the direct transmission.
Motivated by this observation, the RA problem is formulated by defining:
\begin{itemize}
\item
$\tklR\in \{0,1\}$ and $\Pkl\geq 0$, $\forall\;k,l$.
$\tklR=1$ indicates that subcarrier $k$ is paired with subcarrier $l$
for the relay-aided transmission.
When $\tklR=1$, $\Pkl$ is used as the total power for the subcarrier pair $(k,l)$.

\item
$\tklD\in \{0,1\}$, $\pkl\geq 0$ and $\qkl\geq 0$, $\forall\;k,l$.
$\tklD=1$ indicates that subcarriers $k$ and $l$ form a virtual subcarrier pair
for the direct transmission.
When $\tklD=1$,  $\Pskone$ and $\Psltwo$ take the value of $\pkl$ and $\qkl$, respectively.
\end{itemize}

Let us collect all indicator and power variables in the sets
$\Iset$ and $\Pset$, respectively, and define $\RA = \{\Iset,\Pset\}$.
The RA problem can be formulated as the problem (P1):
\begin{align}
\min_{\RA}  &\hspace{0.25cm} \sum_{k,l}(\tklR\Pkl +\tklD\pkl+\tklD\qkl),    \nonumber\\
 {\rm s.t.}  &\hspace{0.25cm}  \tklR,\tklD\in\{0,1\},\forall\;k,l;          \nonumber\\
             &\hspace{0.25cm} \sum_{l}\left(\tklD + \tklR\right) = 1, \forall\;k;
             \hspace{0.1cm}  \sum_{k}\left(\tklD + \tklR\right) = 1, \forall\;l;     \nonumber\\
            & \hspace{0.25cm}  \Pkl\geq 0, \pkl\geq 0, \qkl\geq 0,\forall\;k,l;  \nonumber\\
            &\hspace{0.25cm}  f(\RA)\geq \Rreq,            \nonumber
\end{align}
where $f(\RA)$ represents the maximum sum rate as
\begin{align}
f(\RA) = \sum_{k,l}\big(\tklR \Rate{\Gkl \Pkl} +
          \tklD\Rate{\Gsdk \pkl} + \tklD\Rate{\Gsdl\qkl}\big).  \nonumber
\end{align}

Obviously, (P1) is a nonconvex mixed-integer nonlinear program.
To find the optimum $\RA$, we first relax all indicator variables
to be continuous within $[0,1]$.
Then, we make the change of variables (COV) from $\Pset$ to
$\newPset=\{\newPkl,\newpkl,\newqkl|\forall k,l\}$,
where $\newPkl$, $\newpkl$ and $\newqkl$ satisfy
$\newPkl = \tklR\Pkl$, $\newpkl=\tklD\pkl$ and $\newqkl=\tklD\qkl$, respectively, $\forall\;k,l$.
After collecting all variables into $\newRA = \{\Iset,\newPset\}$,
the RA problem can be rewritten as the problem (P2):
\begin{align}
\min_{\newRA}  &\hspace{0.25cm} \sumP(\newRA)=\sum_{k,l}(\newPkl+\newpkl+\newqkl)    \nonumber\\
{\rm s.t.}     &\hspace{0.25cm}  \tklR,\tklD\in[0,1],\forall\;k,l;          \nonumber\\
               &\hspace{0.25cm} \sum_{l}\left(\tklD + \tklR\right) = 1, \forall\;k;
                 \hspace{0.1cm}  \sum_{k}\left(\tklD + \tklR\right) = 1, \forall\;l;     \nonumber\\
               & \hspace{0.25cm}  \newPkl\geq 0, \newpkl\geq 0, \newqkl\geq 0,\forall\;k,l;  \nonumber\\
               &\hspace{0.25cm}    -g(\newRA)\leq -\Rreq,            \nonumber
\end{align}
where $g(\newRA)$ represents the maximum sum rate expressed as
\begin{align}
g(\newRA) = \sum_{k,l} &\big(\phi(\tklR,\newPkl,\Gkl)                          \nonumber\\
             & + \phi(\tklD,\newpkl,\Gsdk) + \phi(\tklD,\newqkl,\Gsdl)\big),   \nonumber
\end{align}
and
\begin{align}
\phi(t,x,G) = \left\{\begin{array}{ll}
                     t \, \Rate{G \, \frac{x}{t}}   &   {\rm if\;}t>0,  \\
                     0                              &   {\rm if\;}t=0.  \\
                   \end{array}\right.
\end{align}

Obviously (P2) is a relaxation of (P1).
We will find an (at least approximately) optimum solution for (P2),
and show that the $\RA$ corresponding to this solution is still feasible,
and hence (at least approximately) optimum for (P1).
To this end, note that $\phi(t,x,G)$ with fixed $G$
is a continuous and concave function of $t\geq 0$ and $x$,
because it is a perspective function of $\Rate{G x}$ which is concave of $x$ \cite{Convex-opt}.
As a result, $g(\newRA)$ is a concave function of $\newRA$ in its feasible domain for (P2).
This means that (P2) is a convex optimization problem.
As can be checked, it also satisfies the Slater constraint qualification,
therefore it has zero duality gap,
which justifies the applicability of the dual method to find 
the globally optimum for (P2), denoted as $\newRA^\star$ hereafter.

To use the dual method,
$\mu$ is introduced as a Lagrange multiplier for the rate constraint.
The Lagrange relaxation problem for (P2) is the problem (P3):
\begin{align}
\min_{\newRA}  &\hspace{0.25cm} L(\mu,\newRA) = \sumP(\newRA) + \mu\bigg(\Rreq - g(\newRA)\bigg)  \nonumber\\
{\rm s.t.}     &\hspace{0.25cm}  \tklR,\tklD\in[0,1],\forall\;k,l;          \nonumber\\
               &\hspace{0.25cm} \sum_{l}\left(\tklD + \tklR\right) = 1, \forall\;k;
                 \hspace{0.1cm}  \sum_{k}\left(\tklD + \tklR\right) = 1, \forall\;l;     \nonumber\\
               & \hspace{0.25cm}  \newPkl\geq 0, \newpkl\geq 0, \newqkl\geq 0,\forall\;k,l;  \nonumber
\end{align}
where $L(\mu,\newRA)$ is the Lagrangian of (P2).
A global optimum for (P3) is denoted by $\newRA_\mu$
and the dual function is defined as $d(\mu)=L(\mu,\newRA_\mu)$.
Note that $d(\mu)$ is concave of $\mu\geq0$, and $\Rreq - g(\newRA_\mu)$
is a subgradient of $d(\mu)$, i.e.,
$\forall\;\mu'$, $d(\mu') \leq  d(\mu) + (\mu'-\mu)(\Rreq-g(\newRA_\mu))$.
The dual problem is to find the dual optimum $\mu^\star = \arg\min_{\mu\geq0}d(\mu)$.

Since (P2) has zero duality gap, two important properties should be noted.
One is that $\mu^\star>0$. This is because $\mu^\star$ represents the sensitivity
of the optimum objective value for (P2) with respect to $\Rreq$, i.e.,
$\frac{\partial\sumP(\newRA^\star)}{\partial\Rreq} = \mu^\star$ \cite{Convex-opt}.
Obviously, $\sumP(\newRA^\star)$ is strictly increasing of $\Rreq$,
meaning that $\mu^\star>0$.
The other is that $\mu = \mu^\star$ and $\newRA_\mu=\newRA^\star$,
if and only if $\newRA_\mu$ is feasible and $\mu(g(\newRA_\mu)-\Rreq) = 0$
according to Proposition $5.1.5$ in \cite{Nonlinear-opt}.
Based on the above property, the $\mu>0$ and $\newRA_{\mu}$ that satisfies
$g(\newRA_\mu)=\Rreq$ can be found as $\mu^\star$ and $\newRA^\star$.
Therefore, the key to using the dual method consists of
two procedures to finding $\newRA_\mu$ and $\mu^\star$, respectively.
We first introduce the one to finding $\newRA_\mu$ as follows.

\subsubsection{To find $\newRA_\mu$ when $\mu>0$}\label{sec:solve-LRP}
the following strategy is used. First, the optimum $\newPset$ for (P3)
with fixed $\Iset$ is found and denoted by $\newPset_\Iset$.
Define $\newRA_\Iset = \{\Iset, \newPset_\Iset\}$.
Then we find the optimum $\Iset$ to maximize $L(\mu,\newRA_\Iset)$
subject to the constraints on $\Iset$ in (P3).
$\newRA_\Iset$ corresponding to this optimum $\Iset$ can be taken as $\newRA_\mu$.

Suppose $\Iset$ is fixed, it can readily be shown that
the optimum $\newPkl$, $\newpkl$ and $\newqkl$ for (P3) are
\begin{align}
\newPkl = \tklR \Lambda(\mu,\Gkl);
\newpkl = \tklD\Lambda(\mu,\Gsdk);  \newqkl = \tklD\Lambda(\mu,\Gsdl)  \nonumber
\end{align}
where $\Lambda(\mu,G) = \left[\frac{\log_2{e}}{2}\mu - \frac{1}{G}\right]^+$.
Using these formulas, $\newRA_\Iset = \{\Iset, \newPset_\Iset\}$
can be found. It can readily be shown that
\begin{align}
L(\mu,\newRA_\Iset) = \mu\Rreq + \sum_{k,l} \big(\tklR\Akl + \tklD\Bkl\big)
\end{align}
where
\begin{align}
\Akl = &\Lambda(\mu,\Gkl) - \mu\cdot\Rate{\Gkl\Lambda(\mu,\Gkl)}     \nonumber\\
\Bkl = &\Lambda(\mu,\Gsdk) - \mu\cdot\Rate{\Gsdk\Lambda(\mu,\Gsdk)}   \nonumber\\
        &+ \Lambda(\mu,\Gsdl) - \mu\cdot\Rate{\Gsdl\Lambda(\mu,\Gsdl)}    \nonumber.
\end{align}

Now, it can be readily shown that the optimum $\Iset$ for (P3)
is the solution to the problem (P4),
\begin{align}
\min_{\Iset,\{\tkl|\forall\;k,l\}} &\hspace{0.25cm} \sum_{k,l} \big(\tklR\Akl + \tklD\Bkl\big)   \nonumber\\
{\rm s.t.}     &\hspace{0.25cm}  \tklR,\tklD,\tkl\in[0,1],\forall\;k,l;    \nonumber\\
               & \hspace{0.25cm} \tkl = \tklR + \tklD,\forall\;k,l;       \nonumber\\
               &\hspace{0.25cm} \sum_{l}\tkl = 1, \forall\;k;
                \hspace{0.25cm} \sum_{k}\tkl = 1, \forall\;l;             \nonumber
\end{align}
where extra variables $\{\tkl|\forall\;k,l\}$ are introduced.
Note that $\tklR\Akl + \tklD\Bkl \geq \tkl \Ckl$ holds
where $\Ckl = \min\{\Akl, \Bkl\}$.
Let us label $\Akl$ as the metric for $\tklR$ and $\Bkl$ as the metric for $\tklD$.
This inequality is tightened when the entry in $\{\tklR,\tklD\}$
with the smaller metric is assigned to $\tkl$, while the other entry assigned to $0$.
This means that after the problem (P5):
\begin{align}
\min_{\{\tkl|\forall\;k,l\}}  &\hspace{0.25cm} \sum_{k,l} \tkl\Ckl  \nonumber\\
{\rm s.t.}    &\hspace{0.25cm} \tkl \in [0,1], \forall\;k,l;        \nonumber\\
              & \hspace{0.25cm}\sum_{l}\tkl = 1, \forall\;k;
               \hspace{0.25cm} \sum_{k}\tkl = 1, \forall\;l;         \nonumber
\end{align}
is solved for its optimum solution $\{\tkl^\star|\forall\;k,l\}$,
an optimum $\Iset$ for (P4) can be constructed as follows.
For every combination of $k$ and $l$,
the entry in $\{\tklR,\tklD\}$ with the smaller metric is assigned with
$\tkl^\star$, while the other entry with $0$.

Most interestingly, (P5) is a standard assignment problem,
hence $\{\tkl^\star|\forall\;k,l\}$ can be found efficiently by the Hungarian algorithm,
and every entry in $\{\tkl^\star|\forall\;k,l\}$ is either $0$ or $1$ \cite{Hungarian}.
After knowing $\{\tkl^\star|\forall\;k,l\}$, the optimum $\Iset$
can be constructed according to the way mentioned earlier.
Finally, the corresponding $\newRA_\Iset = \{\Iset,\newPset_\Iset\}$ is assigned to $\newRA_\mu$.
Note that the Hungarian algorithm to solve (P5) has a complexity of $O(K^3)$ \cite{Hungarian}.

\subsubsection{To find $\mu^\star$}
an incremental-update based subgradient method can be used as in \cite{Haj11,Hsu10}.
However, this method converges very slowly. 
To develop a faster algorithm, we first show that
$g(\newRA_\mu)$ is a non-decreasing function of $\mu\geq0$.
To this end, suppose $\mu_1\geq \mu_2$.
Since $\Rreq-g(\newRA_{\mu})$ is a subgradient of $d(\mu)$ at $\mu$,
$d(\mu_1) \leq  d(\mu_2) + (\mu_1-\mu_2)(\Rreq-g(\newRA_{\mu_2}))$ and
$d(\mu_2) \leq  d(\mu_1) + (\mu_2-\mu_1)(\Rreq-g(\newRA_{\mu_1}))$
follow. As a result,
\begin{align}
(\mu_1 - \mu_2)(\Rreq- g(\newRA_{\mu_1})) &\leq  d(\mu_1) - d(\mu_2)   \nonumber\\
   &\leq  (\mu_1-\mu_2)(\Rreq- g(\newRA_{\mu_2}))                     \nonumber
\end{align}
holds, and thus $g(\newRA_{\mu_1})) \geq g(\newRA_{\mu_2})$,
meaning that $g(\newRA_\mu)$ is indeed non-decreasing with $\mu$.
Based on the above property, the bisection method can be used to
the $\mu>0$ satisfying $g(\newRA_\mu)=\Rreq$ as $\mu^\star$.

The overall procedure to solving (P2) for $\newRA^\star$ is shown
in Algorithm \ref{alg:dual},
where $\epsilon>0$ is small and prescribed.
As can be shown in a similar way as in \cite{WangTSP13},
the finally produced $\newRA_{\mu}$  is either equal to
(if $g(\newRA_\mu)=\Rreq$ is satisfied),
or a close approximation (if $\Rreq< g(\newRA_\mu)\leq \Rreq+\epsilon$ is satisfied) for $\newRA^\star$.
Moreover, the indicator variables in that $\newRA_{\mu}$
are either $0$ or $1$, and therefore the corresponding $\RA$
is either optimum or approximately optimum for (P1).
It can readily be shown that Algorithm \ref{alg:dual} has
a polynomial complexity with respect to $K$.

\begin{algorithm}
\caption{The algorithm to solve (P1).} \label{alg:dual}
\begin{algorithmic}[1]
\STATE  compute $\Gkl$, $\forall\;k,l$; 
\STATE  $\mumin = 0$; $\mumax=1$; compute $g(\newRA_{\mumax})$;
\WHILE{$g(\newRA_{\mumax})\leq \Rreq$}
       \STATE  $\mumax = 2\mumax$;  compute $g(\newRA_{\mumax})$;
\ENDWHILE

\WHILE{1}
       \STATE  $\mu = \frac{\mumax+\mumin}{2}$;  solve (P3) for $\newRA_\mu$;
       \IF{$\Rreq\leq g(\newRA_\mu)\leq \Rreq+\epsilon$}
           \STATE  go to line 15;
       \ELSIF{$g(\newRA_\mu) > \Rreq+\epsilon$}
           \STATE  $\mumax = \mu$;
       \ELSE
           \STATE  $\mumin = \mu$;
       \ENDIF
\ENDWHILE
\STATE  compute the $\RA$ corresponding to $\newRA_{\mu}$
        as an (at least approximately) optimum solution for (P1).
\end{algorithmic}
\end{algorithm}

\section{Numerical experiments}

Consider the scenario where the relay is located
in the straight line between the source and the destination.
The source-to-destination and source-to-relay distances
are $1$ km and $d$ km ($d\in[0,1]$), respectively.
The parameters are set as $\sigma^2 = -50$ dBm, $\Rreq = 100$ bpos
and $\epsilon = 1$.
When $K$ and $d$ are fixed, every channel impulse response is
randomly generated in the same way as in \cite{Wang11TSP-1}. 

To illustrate the power-reduction benefit of the improved DF protocol with OSP,
two benchmark protocols are considered.
The first one is the improved DF protocol with a priori subcarrier pairing
as studied in \cite{Vandendorpe08-2}.
The second one is the noncooperative transmission,
i.e., the direct transmission is used at every subcarrier.
Define $\PminSP$, $\PminFSP$ and $\PminD$
as the minimum sum power needed for the improved DF protocol with OSP,
the first and second benchmark protocols, respectively.
Define $\Nsp$ and $\Nfsp$ as the optimum number of subcarrier pairs used
for the relay-aided transmission by the improved DF protocol with OSP
and the first benchmark protocol, respectively.
$\PminSP$ and $\Nsp$ can be computed with Algorithm 1.
It can readily be shown that $\PminD = 2\sum_k\left[\lambda - \frac{1}{\Gsdk}\right]^+$,
where $\lambda$ satisfies that $\sum_{k}\Rate{[\lambda\Gsdk-1]^+} = \frac{\Rreq}{2}$.
Moreover, $\PminFSP$ is equal to the optimum objective value of (P1)
imposed with the extra constraint $\tklR=\tklD=0$, $\forall\; k,l:k\neq l$.
An algorithm similar as Algorithm \ref{alg:dual} can be designed
to find $\PminFSP$ and $\Nfsp$, which is omitted here due to space limitation.

\begin{figure}[h]
  \centering
     \includegraphics[width=3.5in,height=2in]{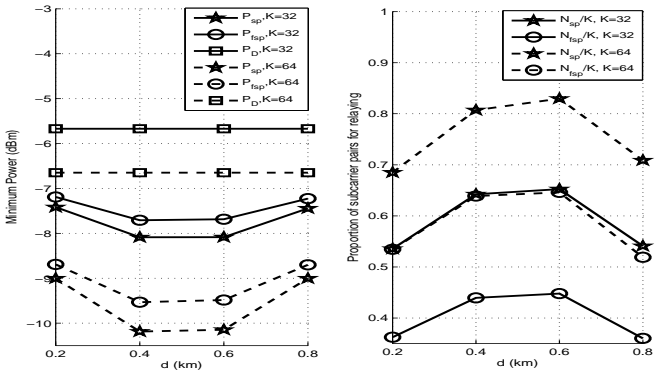}
  \caption{The numerical results for different combinations of $K$ and $d$.}  \label{fig:K-fixed}
\end{figure}

We have computed the average $\PminSP$, $\PminFSP$, $\PminD$,
$\frac{\Nsp}{K}$ and $\frac{\Nfsp}{K}$ over $1000$ random channel realizations
for different combinations of $K$ and $d$.
The results are shown in Figure \ref{fig:K-fixed}.
It is shown that for any fixed combination of $K$ and $d$,
the average $\PminSP$ is smaller than the average $\PminFSP$ and $\PminD$,
which illustrates the power-reduction benefit of the improved DF protocol with OSP.
Moreover, the average $\PminFSP$ is smaller than the average $\PminD$.
This is because the first benchmark protocol uses opportunistic DF relaying,
which better exploits the flexibility of transmission-mode selection
for the sum-power reduction.

When $K$ is fixed, it can be seen that
the average $\PminSP$ and $\PminFSP$ reduce
while the average $\frac{\Nsp}{K}$ and $\frac{\Nfsp}{K}$ increase
if the relay moves towards the middle between the source and the relay.
This trend for the average $\PminSP$ and $\frac{\Nsp}{K}$
is explained as follows
(the one for the average $\PminFSP$ and $\frac{\Nfsp}{K}$
can be explained in a similar way).
Obviously, the pairing of more subcarriers for the relay-aided transmission
is more beneficial for sum-power reduction
if $\forall\;k,l$, $\Gkl$ is more likely to take a high value.
Note that $\Gkl$ takes a high value
only if both $\Gsrk$ and $\Grdl$ are much higher than $\Gsdk$,
which can be verified by using the intuitive method
explained in the Appendix of \cite{WangTSP11}.
When the relay lies in the middle between the source and the relay,
it is more likely to have $\Gsrk$ and $\Grdl$ both be much greater than $\Gsdk$,
and thus $\Gkl$ is more likely to take a high value.
This explains the observation.

When $d$ is fixed and $K$ increases,
it can be observed that the average $\PminSP$ and $\PminFSP$ reduce
while the average $\frac{\Nsp}{K}$ and $\frac{\Nfsp}{K}$ increase.
Moreover, the average $\PminSP$ and $\PminFSP$ are much smaller than $\PminD$,
and the average $\PminSP$ is much smaller than the average $\PminFSP$,
especially when $K$ takes a high value.
This is because using more subcarriers leads to
more flexibility of subcarrier pairing and transmission-mode selection
for the sum-power reduction.

\section{Conclusion}

We have developed a sum-power minimized RA algorithm
subject to a sum-rate constraint for cooperative OFDM transmission
using the improved DF protocol with optimized subcarrier pairing.
The power-reduction benefit of this protocol
has been illustrated by numerical results.



\ifCLASSOPTIONcaptionsoff
  \newpage
\fi



\bibliographystyle{IEEEtran}
\bibliography{subcarr-pairing}%

\begin{thebibliography}{10}
\providecommand{\url}[1]{#1}
\csname url@samestyle\endcsname
\providecommand{\newblock}{\relax}
\providecommand{\bibinfo}[2]{#2}
\providecommand{\BIBentrySTDinterwordspacing}{\spaceskip=0pt\relax}
\providecommand{\BIBentryALTinterwordstretchfactor}{4}
\providecommand{\BIBentryALTinterwordspacing}{\spaceskip=\fontdimen2\font plus
\BIBentryALTinterwordstretchfactor\fontdimen3\font minus
  \fontdimen4\font\relax}
\providecommand{\BIBforeignlanguage}[2]{{%
\expandafter\ifx\csname l@#1\endcsname\relax
\typeout{** WARNING: IEEEtran.bst: No hyphenation pattern has been}%
\typeout{** loaded for the language `#1'. Using the pattern for}%
\typeout{** the default language instead.}%
\else
\language=\csname l@#1\endcsname
\fi
#2}}
\providecommand{\BIBdecl}{\relax}
\BIBdecl

\bibitem{Vandendorpe08-2}
L.~Vandendorpe, J.~Louveaux, O.~Oguz \emph{et~al.}, ``Improved {OFDM}
  transmission with {DF} relaying and power allocation for a sum power
  constraint,'' in \emph{ISWPC}, 2008, pp. 665--669.

\bibitem{WangYing07}
Y.~Wang, X.~Qu, T.~Wu, and B.~Liu, ``Power allocation and subcarrier pairing
  algorithm for regenerative ofdm relay system,'' in \emph{IEEE Veh. Techn.
  Conf.}, Apr. 2007, pp. 2727--2731.

\bibitem{Li08}
Y.~Li, W.~Wang, J.~Kong \emph{et~al.}, ``Power allocation and subcarrier
  pairing in {OFDM}-based relaying networks,'' in \emph{IEEE Int. Conf.
  Commun.}, May 2008, pp. 2602--2606.

\bibitem{Haj11}
M.~Hajiaghayi, M.~Dong, and B.~Liang, ``Optimal channel assignment and power
  allocation for dual-hop multi-channel multi-user relaying,'' in \emph{Proc.
  IEEE INFOCOM}, Apr. 2011, pp. 76 --80.

\bibitem{Vandendorpe08-1}
L.~Vandendorpe, R.~Duran, J.~Louveaux \emph{et~al.}, ``Power allocation for
  {OFDM} transmission with {DF} relaying,'' in \emph{IEEE Int. Conf. Commun.},
  May 2008, pp. 3795--3800.

\bibitem{Vandendorpe09-1}
L.~Vandendorpe, J.~Louveaux, O.~Oguz \emph{et~al.}, ``Power allocation for
  improved {DF} relayed {OFDM} transmission: the individual power constraint
  case,'' in \emph{IEEE Int. Conf. Commun.}, 2009, pp. 1--6.

\bibitem{Vandendorpe09-2}
------, ``Rate-optimized power allocation for {DF}-relayed {OFDM} transmission
  under sum and individual power constraints,'' \emph{Eurasip J. on Wirel.
  Commun. and Networking}, vol. 2009.

\bibitem{WangTSP11}
T.~Wang and L.~Vandendorpe, ``{WSR} maximized resource allocation in multiple
  {DF} relays aided {OFDMA} downlink transmission,'' \emph{IEEE Trans. Sig.
  Proc.}, vol.~59, no.~8, pp. 3964 --3976, Aug. 2011.

\bibitem{WangJSAC11}
------, ``Sum rate maximized resource allocation in multiple {DF} relays aided
  {OFDM} transmission,'' \emph{IEEE J. Sel. Areas on Commun.}, vol.~29, no.~8,
  pp. 1559 --1571, Sep. 2011.

\bibitem{Hsu10}
C.-N. Hsu, P.-H. Lin, and H.-J. Su, ``Joint subcarrier pairing and power
  allocation for ofdm two-hop systems,'' in \emph{ICC 2010}, May 2010, pp. 1
  --5.

\bibitem{Boost10}
H.~Boostanimehr, O.~Duval, V.~Bhargava, and F.~Gagnon, ``Selective subcarrier
  pairing and power allocation for decode-and-forward ofdm relay systems,'' in
  \emph{ICC 2010}, May 2010, pp. 1 --5.

\bibitem{Li11}
G.~Li, Z.~Xu, C.~Xiong, C.~Yang, S.~Zhang, Y.~Chen, and S.~Xu,
  ``Energy-efficient wireless communications: tutorial, survey, and open
  issues,'' \emph{IEEE Wireless Communications}, vol.~18, no.~6, pp. 28 --35,
  Dec. 2011.

\bibitem{Convex-opt}
S.~Boyd and L.~Vandenberghe, \emph{Convex optimization}.\hskip 1em plus 0.5em
  minus 0.4em\relax Cambridge University Press, 2004.

\bibitem{Nonlinear-opt}
D.~P. Bertsekas, \emph{Nonlinear programming, $2$nd edition}.\hskip 1em plus
  0.5em minus 0.4em\relax Athena Scientific, 2003.

\bibitem{Hungarian}
H.~Khun, ``The hungarian method for the assignment problems,'' \emph{Naval
  Research Logistics Quarterly 2}, pp. 83--97, 1955.

\bibitem{WangTSP13}
T.~Wang, F.~Glineur, J.~Louveaux, and L.~Vandendorpe, ``{WSR} maximization for
  downlink {OFDMA} with subcarrier-pair based opportunistic {DF} relaying,''
  \emph{submitted to IEEE Trans. Signal Proc.}, 2013.

\bibitem{Wang11TSP-1}
T.~Wang and L.~Vandendorpe, ``Iterative resource allocation for maximizing
  weighted sum min-rate in downlink cellular {OFDMA} systems,'' \emph{IEEE
  Trans. Signal Process.}, vol.~59, no.~1, pp. 223--234, Jan. 2011.

\end{thebibliography}



\end{document}